\pdfoutput=1
\PassOptionsToPackage{sort&compress,numbers,super}{natbib}

\documentclass[reprint]{revtex4-1} 
\usepackage{achemso}
\usepackage[english]{babel}
\usepackage{graphicx}
\usepackage{amsmath}
\usepackage{amsfonts}
\usepackage{color}
\usepackage{bm}
\usepackage{units}
\usepackage{siunitx}
\usepackage{placeins}
\usepackage[final]{changes}
\begin{document}

\title{Anisotropic Nonequilibrium Lattice Dynamics of Black Phosphorus}

\author{Daniela Zahn}
%\email{zahn@fhi-berlin.mpg.de}
\author{Patrick-Nigel Hildebrandt}
\author{Thomas Vasileiadis}
\author{\mbox{Yoav William Windsor}}
\author{Yingpeng Qi}
\author{H\'{e}l\`{e}ne Seiler}
\author{Ralph Ernstorfer}
%\email{ernstorfer@fhi-berlin.mpg.de}
\affiliation{Fritz Haber Institute of the Max Planck Society, Faradayweg 4-6, 14195 Berlin, Germany}

\begin{abstract}
Black phosphorus has recently attracted significant attention for its highly anisotropic properties. A variety of ultrafast optical spectroscopies has been applied to probe the carrier response to photoexcitation, but the complementary lattice response has remained unaddressed. Here we employ femtosecond electron diffraction to explore how the structural anisotropy impacts the lattice dynamics after photoexcitation. We observe two time scales in the lattice response, which we attribute to electron-phonon and phonon-phonon thermalization. Pronounced differences between armchair and zigzag directions are observed, indicating a nonthermal state of the lattice lasting up to {\raise.17ex\hbox{$\scriptstyle\sim$}}\hspace{2pt}60\hspace{2pt}ps. This nonthermal state is characterized by a modified anisotropy of the atomic vibrations compared to equilibrium. Our findings provide insights in both electron-phonon as well as phonon-phonon coupling and bear direct relevance for any application of black phosphorus in nonequilibrium conditions.
\end{abstract}  
\maketitle

\begin{figure}[b!]
   \includegraphics[width=\columnwidth]{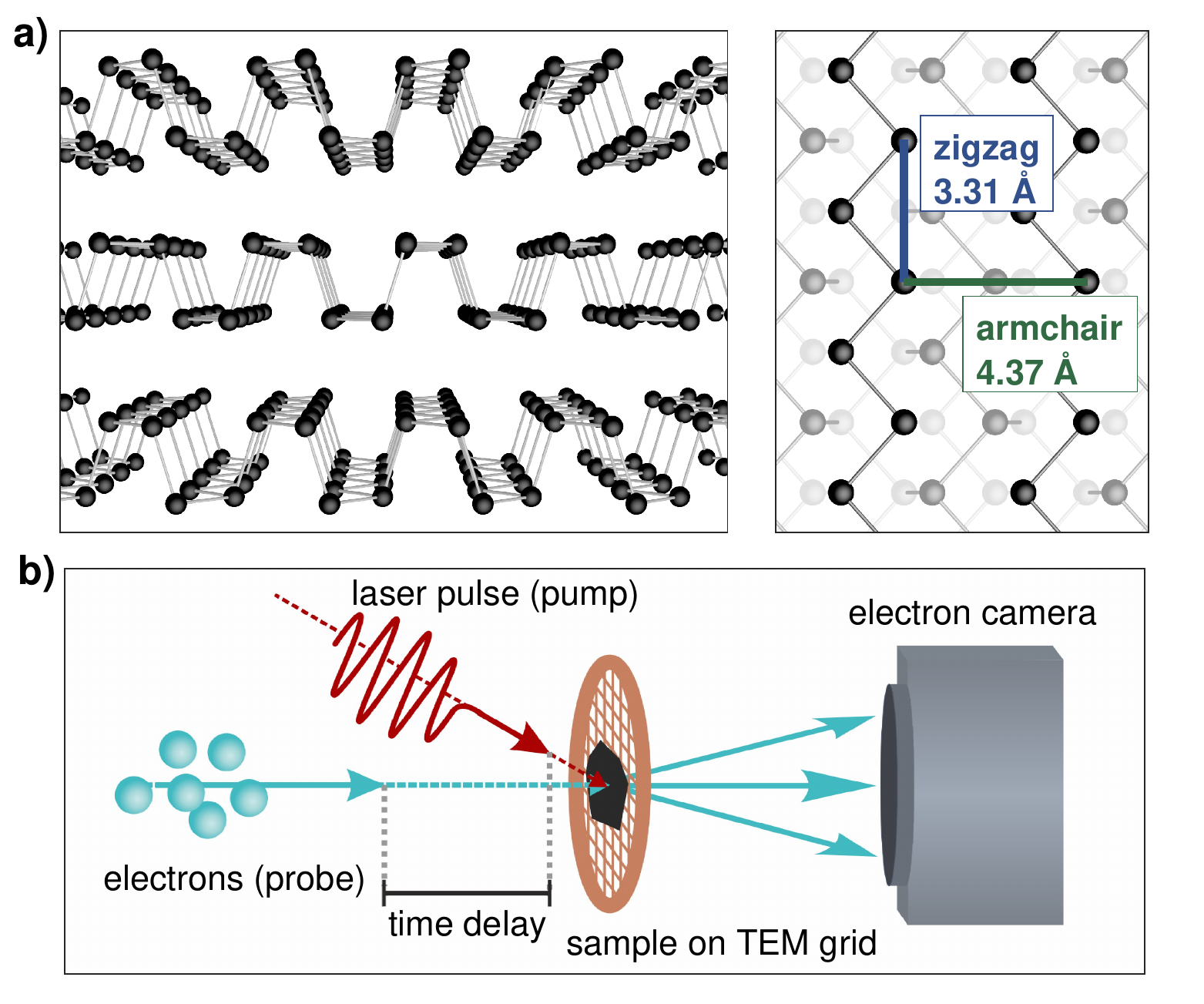}
  \caption{Anisotropic structure of black phosphorus and schematic illustration of the experiment. (a) Side and top views of the crystal structure of black phosphorus, showing anisotropy not only between the in-plane and out-of-plane directions but also between the in-plane directions (armchair and zigzag). 
  (b) Schematic representation of a time-resolved electron diffraction setup (see text for details).}
  \label{fig:1}
\end{figure}

Layered van der Waals (vdW) materials have attracted significant research interest in recent years due to their potential device applications \cite{2015Lotsch,2016Novo,2017Schulman,2019Cheng}. The most prominent 2D material, graphene, exhibits high carrier mobility, but lacks a band gap, which is required in many applications. In contrast, transition metal dichalcogenides possess a band gap in the visible range, but a lower carrier mobility. With a thickness-dependent band gap extending from the infrared to the visible \cite{2014Qiao,2015Cast,2016Li} and a high carrier mobility \cite{2014Li,2014Xia,2016Long}, black phosphorus provides an important complementary building block for vdW heterostructure devices. A central aspect of black phosphorus is its in-plane anisotropic structure, shown in Figure \ref{fig:1}a. The layers have two inequivalent high-symmetry directions, the so-called zigzag and armchair directions. This structural anisotropy is also reflected in many macroscopic material properties, such as optical absorption \cite{2014Xia,2014Tran,2014Low,Lan2016,2018Jiang} and in-plane anisotropic thermal \cite{2015Lee,2015Luo,2015Jang,2017Sun} and electrical  \cite{2014Liu,2014Xia,2014Qiao,2015He} conductivities. These anisotropic properties offer additional tunability in device design.

Since any device operates in nonequilibrium conditions, a microscopic understanding of nonequilibrium states in vdW materials is of particular interest. For optoelectronic devices, knowledge of the evolution of the system after optical excitation is desired. Carrier dynamics in black phosphorus have been studied using a variety of time-resolved optical spectroscopies \cite{2015He,2015Ge,2015Suess,2016Wang,2017Iyer,2017Liao,2019Meng} as well as time- and angle-resolved photoemission \cite{2019Roth,2019Chen} (trARPES). An important relaxation pathway for excited carriers is via coupling to the lattice. However, to date, no study has directly reported on the ultrafast lattice response of black phosphorus upon photoexcitation, which reflects the strength of electron-phonon as well as phonon-phonon interactions. In this work, we employ femtosecond electron diffraction \cite{2015Wald} (FED) to directly probe the structural dynamics of photoexcited black phosphorus. 

%%%%%%%%%%%%%%%% FIGURE 1 %%%%%%%%%%%%%%%%%%%%%%%%%%%%%%%%%%%%%
%%%%%%%%%%%%%%%%%%%%%%%%%%%%%%%%%%%%%%%%%%%%%%%%%%%%%%%%%%%%%%% 
The measurement principle is sketched in Figure \ref{fig:1}b. The sample is excited with an ultrashort laser pulse (pump) and the lattice response is probed using an ultrashort electron pulse (probe) with a kinetic energy of \unit[70]{keV}. The electrons are diffracted by the sample and diffraction patterns are recorded in transmission for different time delays between pump and probe pulses. To excite the sample, we use optical pulses with a wavelength of \unit[770]{nm} (\unit[1.61]{eV}). The polarization of the pump pulse is set to the armchair direction of the crystal.
All measurements are performed at a base temperature of \unit[100]{K}. 

Since diffraction patterns are measured in transmission, the samples need to be thin films. We prepared a thin film of black phosphorus by mechanical exfoliation from a bulk crystal (\textit{HQ Graphene}) using water-soluble glue and transferred it on a standard copper TEM grid using the floating technique \cite{2007Dwyer}. Based on the transmission of the film and previously reported optical constants of black phosphorus \cite{2018Jiang}, we estimate the film thickness to be \unit[$39\pm5$]{nm}. The sample was transferred to vacuum directly after preparation to minimize degradation.

%%%%%%%%%%%%%%%% FIGURE 2 %%%%%%%%%%%%%%%%%%%%%%%%%%%%%%%%%%%%%%%%%%
\begin{figure*}[bth!]
   \includegraphics[scale=0.77]{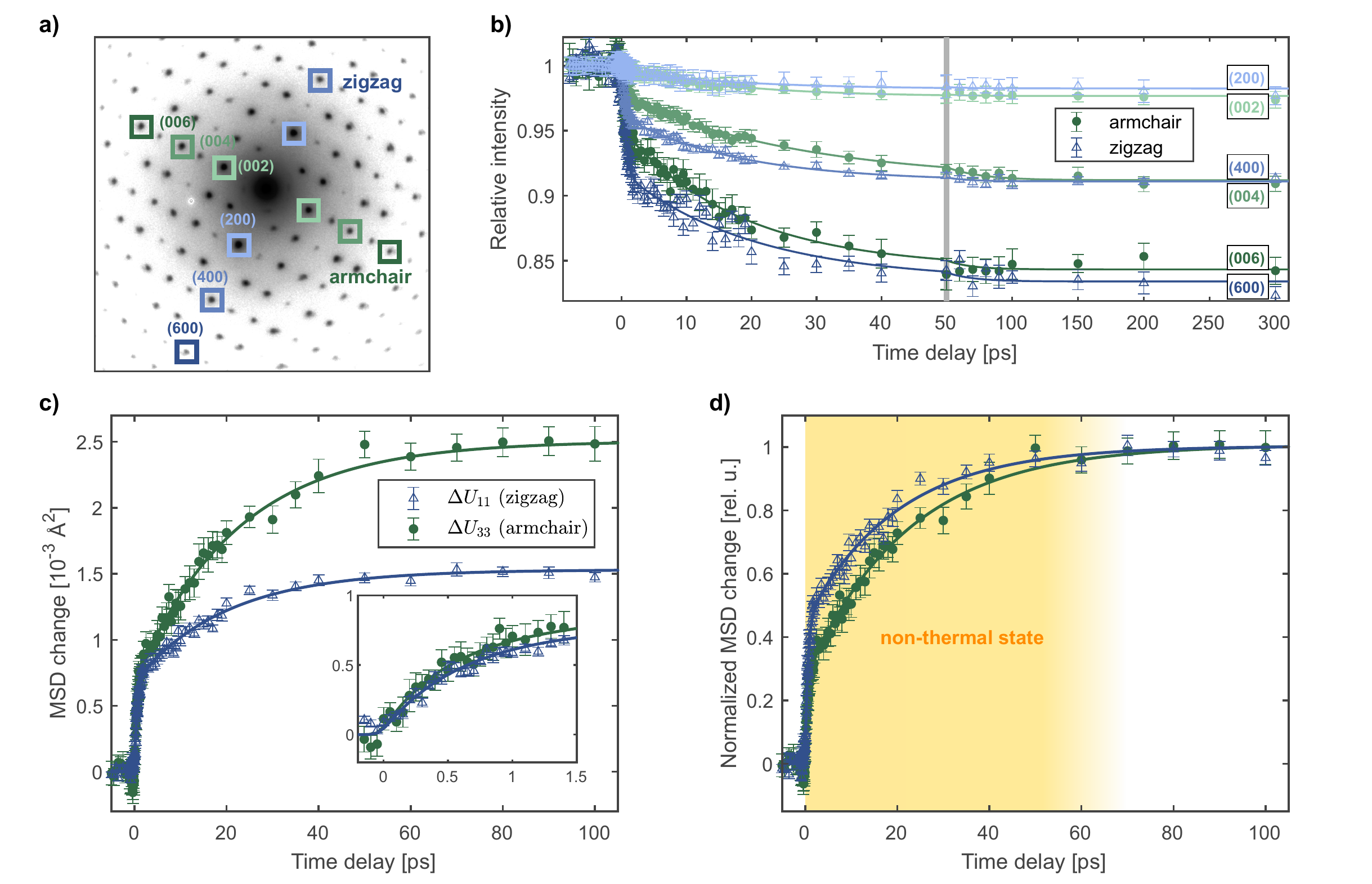}
  \caption{Overview of anisotropic lattice dynamics in photoexcited black phosphorus. (a) Transmission diffraction pattern of thin-film black phosphorus. We focus on the high-symmetry Bragg reflections along the armchair and zigzag directions, indicated by colored boxes. (b) Relative changes in Bragg reflection intensities functions of pump-probe delay. Here, we average over the Friedel pairs, e.g. (002) and (00$\overline{2}$), since they show the same dynamics.  The measurement was conducted with an incident fluence of \unit[($9.8\pm1.4$)]{mJ/cm$^2$}. Based on the optical constants of black phosphorus \cite{2018Jiang} and the film thickness, we estimate the absorbed energy density to be \unit[($380\pm70$)]{J/cm$^3$}. The data presented are the average of several delay scans, and the error estimates represent the standard error of the mean. \mbox{(c) Changes} of atomic mean squared displacement (MSD) in the armchair (green circles) and zigzag (blue triangles) directions as function of pump-probe delay. The anisotropy of the lattice is reflected in an anisotropic MSD change in the two directions. The higher MSD change in the armchair direction indicates that bonds are softer in this direction. The MSD values presented are the weighted average of MSDs calculated from each Friedel pair and the error bars are calculated using error propagation. The inset is a close-up of the data at early time delays. \mbox{(d) MSD change} normalized to the fit value at \unit[100]{ps}. A two-step time scale as well as a transient nonequilibrium between the zigzag and armchair directions is observed.}
  \label{fig:2}
\end{figure*}
%%%%%%%%%%%%%%%%%%%%%%%%%%%%%%%%%%%%%%%%%%%%%%%
Figure \ref{fig:2}a shows a typical transmission diffraction pattern of black phosphorus. The orthorhombic crystal structure of black phosphorus (see Figure \ref{fig:1}a) results in an anisotropic diffraction pattern. Bragg reflections along the zigzag and armchair directions, (h00) and (00l), are marked with blue and green boxes, respectively. Only reflections with even h and l are allowed due to the crystal symmetry (space group Cmce with phosphorus atoms at Wyckoff positions 8f). We observe additional “forbidden” reflections caused by stacking faults, multiple scattering, or structural deviations at the surfaces, which are not taken into account in the following analysis. 

In this work, we focus on the Bragg reflections along the armchair and zigzag directions. Our primary observables are the intensities of the Bragg reflections, which decrease with increasing displacement of the atoms due to lattice vibrations (Debye-Waller effect). Since the probe electrons propagate in parallel to the van der Waals stacking direction through the crystal, our measurement is sensitive to the in-plane atomic vibrations. 

To extract intensities from the diffraction patterns, we fit the observed peaks to the sum of a 2D pseudo-Voigt profile and a tilted background. The same Gaussian-Lorenzian mixing is assumed in all directions. The integrated intensity of the peak is then used to obtain the intensity change as a function of pump-probe delay. Figure \ref{fig:2}b displays the resulting evolution of the Bragg reflection intensities after photoexcitation. We observe two time scales in the intensity decrease and pronounced differences between reflections along the armchair and zigzag directions. The amplitudes of the intensity decrease are larger for reflections with higher scattering vectors, as expected by Debye-Waller theory. 

To analyze the structural dynamics, we convert the Bragg reflection intensities into changes in atomic mean squared displacement (MSD). For anisotropic crystals, the temperature factor is \cite{1996Trueblood,Giacovazzo,Peng}
\begin{equation}
\begin{split}
    \tau=\mathrm{exp}\{-\frac{1}{2} \lbrack U_{11}(ha^*)^2+U_{22}(kb^*)^2+U_{33}(lc^*)^2+\\
+2U_{12}ha^*kb^*+2U_{23}kb^*lc^*+2U_{13}ha^*lc^*\rbrack\}\\
    \end{split}
    \label{DW_aniso_full}
\end{equation}
Here, $h$,$k$ and $l$ are the Miller indices and $a^*$,$b^*$ and $c^*$ are the magnitudes of the reciprocal space lattice vectors, defined such that ${\bf a}\cdot{\bf a}^*=2\pi$. For black phosphorus in the standard setting of Cmce, ${\bf a,b,c}$ are the (real space) lattice vectors in zigzag (a=\unit[3.31]{\AA}), out-of-plane (b=\unit[10.46]{\AA}) and armchair (c=\unit[4.37]{\AA}) directions \cite{1989Akai}. $U_{11}$,$U_{22}$ and $U_{33}$ correspond to the MSDs in zigzag, out-of-plane, and armchair direction, respectively (see references \cite{1996Trueblood,Giacovazzo} for more details about the U-matrix). Note that in black phosphorus, $U_{12}$ and $U_{13}$ are zero due to symmetry \cite{1989Akai,Willis}. For Bragg reflections purely along the zigzag direction, that is (h00), the relative intensity change after laser excitation is given by:
\begin{equation}
\frac{I(t)}{I_0}=\mathrm{exp}\{-[U_{11}(t)-U_{11}^0]\frac{4\pi^2}{a^2}h^2\}
\label{eq:Debye-Waller_zig}
\end{equation}
Similarly, for reflections purely along the armchair direction, (00l), the relative intensity change after laser excitation reads
\begin{equation}
\frac{I(t)}{I_0}=\mathrm{exp}\{-[U_{33}(t)-U_{33}^0]\frac{4\pi^2}{c^2}l^2\}
\label{eq:Debye-Waller_arm}
\end{equation}
Here, $U_{ii}^0$ denotes the MSD before laser excitation and $U_{ii}(t)$ denotes the MSD as a function of pump-probe delay. $I(t)$ denotes the intensity as a function of pump-probe delay and $I_0$ is the intensity before laser excitation. 

The changes in MSD, shown in Figure \ref{fig:2}c, are markedly different for the armchair and zigzag directions. From the different amplitudes we conclude that the interatomic potential is anisotropic. 
It is energetically less costly to displace atoms along the armchair direction compared to the zigzag direction. Hence, as the lattice temperature rises, the additional energy leads to a larger increase of the MSD in the armchair direction. These findings are in qualitative agreement with lattice dynamical calculations \cite{1986Kaneta} and static X-ray diffraction measurements \cite{1989Akai}.

The MSD dynamics in the armchair and zigzag directions are fitted with a biexponential function, see solid lines in Figure \ref{fig:2}c. The finite time resolution is taken into account by convolving the fit function with a Gaussian with a FWHM of \SI{150}{fs}. The fit results for the amplitudes $A_i$ and time constants $\tau_i$ are summarized in \mbox{Table \ref{tab:1}}.
\begin{table}[bth]
    \centering
    \begin{tabular}{|l||l|l|}
    \hline                                          
                                                    &armchair               &zigzag \\
    \hline\hline
            $A_1$ $[10^{-3} \si{\angstrom}^2]$      &$0.66\pm 0.03$        &$ 0.68 \pm 0.02$\\
            $\tau_1 $ [ps]                          &$0.48 \pm 0.05$          &$0.58 \pm 0.04$\\
            $A_2$ $[10^{-3} \si{\angstrom}^2]$      &$1.84 \pm 0.03$        &$0.85 \pm 0.03$\\
            $\tau_2$ [ps]                           &$22\pm 1$              &$20 \pm 2$\\
            \hline
    \end{tabular}
    \caption{Fit results of the MSD in the armchair und zigzag directions with a biexponential function convolved with a Gaussian. The errors correspond to \unit[68.3]{\%} confidence intervals of the fit.}
    \label{tab:1}
\end{table}

The fast time constants $\tau_1$ are very similar for the armchair and zigzag directions. We attribute this initial rise in MSD to energy transfer from the photoexcited electrons to the lattice. The observed \unit[{\raise.17ex\hbox{$\scriptstyle\sim$}}\hspace{2pt}0.5]{ps} electron-lattice equilibration time constant is consistent with subpicosecond dynamics observed with time-resolved optical spectroscopy \cite{2016Wang,2017Iyer,2019Meng} and trARPES \cite{2019Chen}.

While the amplitudes of the initial MSD rise are the same for the two directions, the thermalized state at late delays exhibits a higher MSD increase in the armchair direction compared to the zigzag direction. This indicates that electron-phonon equilibration leads to a nonthermal phonon distribution. Compared to the thermal case, this nonthermal phonon distribution is characterized by a higher ratio of the zigzag to armchair MSD. Hence, on average, phonons with a high displacement in zigzag direction couple more strongly to the electrons. 
The persistence of the nonthermal phonon distribution is illustrated in Figure \ref{fig:2}d,  in which the time-dependent MSDs in the armchair and zigzag directions are normalized to the respective fit values at \unit[100]{ps}. Within tens of picoseconds, the nonthermal phonon distribution relaxes into a thermal phonon distribution. We therefore attribute the slower time constants $\tau_2$ of the MSD dynamics to these redistribution processes. The fact that the MSD rises further during phonon thermalization indicates that high-energy phonons decay into lower-energy phonons, because one high-energy phonon decays into multiple low-energy phonons and in addition, low-energy phonons produce a higher atomic displacement per phonon \cite{Peng}. 

Since the conduction band minimum and the valence band maximum are located at the Z-point of the Brillouin zone \cite{2014Qiao}, thermalized carriers can mostly absorb and emit phonons with small momenta. Hence, we can conclude that lattice thermalization is achieved mostly by direct phonon-phonon coupling via anharmonicities. Finally, a thermal state is reached after \mbox{{\raise.17ex\hbox{$\scriptstyle\sim$}}\hspace{2pt}\unit[60]{ps}}. This time scale of phonon thermalization is similar to time scales reported for the vdW materials WSe$_2$ \cite{2017Wald} and graphite \cite{2018Stern}. 
In contrast to these materials, however, the nonthermal phonon population in black phosphorus manifests itself in an anisotropic evolution of the MSD due to the in-plane anisotropy.

We performed similar experiments for different laser pump fluences, pump pulse polarizations, and for a different sample base temperature. In all cases, we obtain values of {\raise.17ex\hbox{$\scriptstyle\sim$}}\hspace{2pt}\unit[0.5]{ps} for the fast time constants $\tau_1$, which reflect the energy transfer from the electrons to the lattice. 
In contrast, the slow time constants $\tau_2$ decrease with increasing sample base temperature. We attribute this to enhanced phonon-phonon scattering due to a larger phonon population at higher temperature. Correspondingly, we also observe an acceleration of the phonon thermalization with increasing excitation density. For a sample base temperature of \unit[295]{K} and a laser pump fluence of {\raise.17ex\hbox{$\scriptstyle\sim$}}\hspace{2pt}\unit[2]{mJ/cm$^2$}, we obtain $\tau_2$ values of \unit[(11.5$\pm$0.7)]{ps} and \unit[(8$\pm$1)]{ps} for the armchair and zigzag directions, respectively. These data, in addition to the low-temperature data presented here, are available on a data repository \cite{Zenodo}. 

\begin{figure}[hbt!]
   \includegraphics[width=\columnwidth]{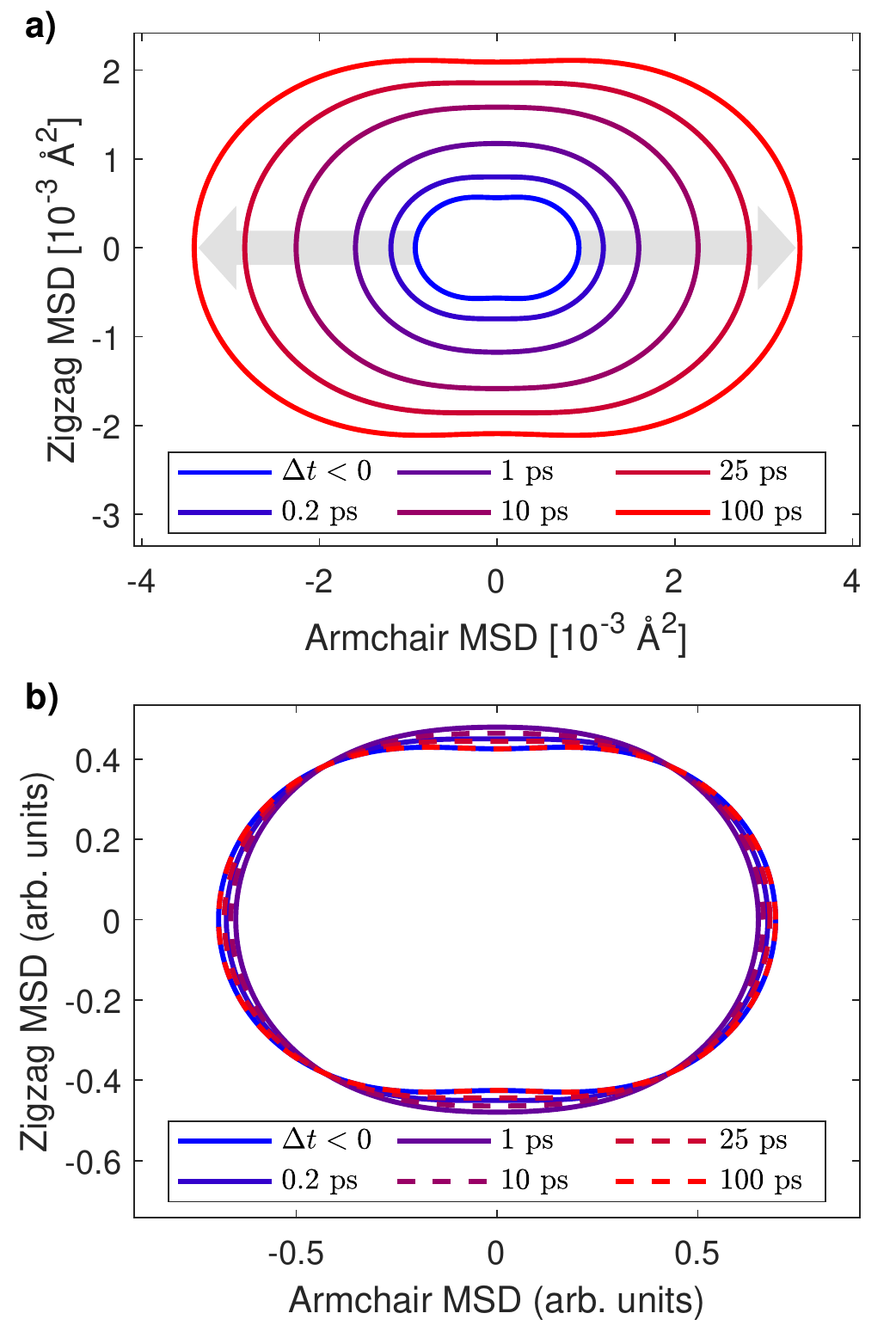}
  \caption{Photoexcitation transiently modifies the anisotropy of the atomic mean squared displacement (MSD). (a) Evolution of the MSD from before excitation (blue) to the thermalized state at \unit[100]{ps} (red). The shape of the in-plane MSD before laser excitation is already anisotropic due to the in-plane structural anisotropy. Note that the MSD increases due to lattice heating but also the shape of the MSD changes transiently. (b) MSD curves from \mbox{panel a}, normalized to their area. This highlights the transient shape change of the MSD due to a nonthermal phonon distribution.}
  \label{fig:4}
\end{figure}

So far, we have only considered the MSD along the armchair and zigzag directions. However, we can calculate the MSD along all in-plane directions based on the MSD along these two directions \cite{Willis,1996Trueblood}. \mbox{Figure \ref{fig:4}} visualizes the evolution of the in-plane MSD, using the biexponential fit of the data shown in \mbox{Figure \ref{fig:2}c}. To estimate the MSD before laser excitation, we make the assumption that both before and after laser excitation, the MSD is proportional to the temperature (high-temperature limit). We estimate that the temperature of the system rises by \unit[($270\pm50$)]{K} after laser excitation, based on the calculated absorbed energy density and the heat capacity of black phosphorus. The total heat capacity is approximated by its main contribution, the lattice heat capacity, calculated from the vibrational DOS \cite{Poseuille}. 
%%%%%%%%%%%%%%%% FIGURE 4 %%%%%%%%%%%%%%%%%%%%%%%%%%%%%%%%%%%%%%%%%%
%%%%%%%%%%%%%%%%%%%%%%%%%%%%%%%%%%%%%%%%%%%%%%%%%%%%%%%

In equilibrium, the shape of the in-plane MSD is already anisotropic (blue curves of Figure \ref{fig:4}) due to the anisotropic bond stiffness. After laser excitation, the MSD increases as shown in Figure \ref{fig:2}c. The evolution of the MSD is displayed in Figure \ref{fig:4}a. Note that not only the size but also the shape of the MSD evolves with time. This is visualized in Figure \ref{fig:4}b by showing the MSD curves normalized to their area. The nonthermal phonon distribution leads to a transient reduction of the MSD anisotropy, with a larger MSD in zigzag direction compared to equilibrium. On longer time scales, as the phonons thermalize, the MSD relaxes back to its equilibrium shape (dashed red curve).

In summary, our measurements have explored how structural anisotropy impacts lattice thermalization in photoexcited black phosphorus. We have shown that the lattice response is well captured by biexponential dynamics: a sub-ps component, common to both zigzag and armchair Bragg reflections, is assigned to electron-phonon equilibration, while a {\raise.17ex\hbox{$\scriptstyle\sim$}}\hspace{2pt}\unit[20]{ps} component is found to be highly anisotropic and indicative of a nonthermal phonon population persisting for {\raise.17ex\hbox{$\scriptstyle\sim$}}\hspace{2pt}\unit[60]{ps}. Our analysis reveals that the nonthermal phonon population results in a transient shape change of the MSD, with a higher displacement in zigzag direction compared to equilibrium.

We expect this nonthermal state of the lattice to have effects on the macroscopic properties of black phosphorus, such as thermal and electrical conductivities. In particular, for any application in which hot carriers are excited or injected in black phosphorus, transient changes of material properties will influence device performance. Beyond black phosphorus, we expect these results to be relevant to other vdW materials as well. For example, in any layered material, the evolution of in-plane and out-of-plane MSD could be different, with implications for energy flow across heterostructure interfaces.  
\newpage
\section*{Acknowledgement}
This work was funded by the Max Planck Society and the European Research Council (ERC) under the European Union’s Horizon 2020 research and innovation program (Grant Agreement Number ERC-2015-CoG-682843). H.S.~acknowledges support by the Swiss National Science Foundation under Grant No.~P2SKP2\textunderscore184100. Y.Q.~acknowledges support by the Sino-German (CSC-DAAD) Postdoc Scholarship Program (Grant No.~57343410).

\bibliography{main.bbl}

\end{document}